\documentclass[sigconf,authorversion,nonacm]{acmart}
\AtBeginDocument{%
  }

\copyrightyear{2025}
\acmYear{2025}
\setcopyright{acmlicensed}\acmConference[CHI'25]{CHI Conference on Human Factors in Computing Systems}{April 26-May 1, 2025}{Yokohama, Japan}
\acmBooktitle{CHI Conference on Human Factors in Computing Systems (CHI'25), April 26-May 1, 2025, Yokohama, Japan}
\acmDOI{10.1145/3706598.3714239}
\acmISBN{979-8-4007-1394-1/25/04}

\usepackage{acmart-taps}
\usepackage{makecell}
\usepackage{soul}
\usepackage{xcolor}
\usepackage{multirow}
\usepackage{listings}
\usepackage{hyperref}
\usepackage{framed}

\definecolor{shadecolor}{rgb}{0.90, 0.95, .98}

\definecolor{dionecolor}{RGB}{36,175,255}
\definecolor{ditwocolor}{RGB}{0,106,183}
\definecolor{caonecolor}{RGB}{255,144,116}
\definecolor{catwocolor}{RGB}{203,31,3}
\definecolor{addtextcolor}{RGB}{195, 38, 38}
\definecolor{camtextcolor}{RGB}{0,100,0}
\definecolor{scenariocolor}{RGB}{230, 244, 252} 
\definecolor{codebgcolor}{gray}{0.95} 
\definecolor{py-propertycolor}{RGB}{0, 85, 170}
\definecolor{py-stringcolor}{RGB}{186, 33, 33}
\definecolor{py-opcolor}{RGB}{170, 34, 255}
\definecolor{py-bracketcolor}{RGB}{153, 153, 119}
\definecolor{py-keywordcolor}{RGB}{0, 128, 0}

\newcommand{\dione}[1]{\textbf{\textcolor{dionecolor}{#1}}}
\newcommand{\ditwo}[1]{\textbf{\textcolor{ditwocolor}{#1}}}
\newcommand{\caone}[1]{\textbf{\textcolor{caonecolor}{#1}}}
\newcommand{\catwo}[1]{\textbf{\textcolor{catwocolor}{#1}}}

\newcommand{\mycode}[1]{\colorbox{codebgcolor}{#1}}
\def\DIFadd#1{#1}
\def\DIFdel#1{}
\def\CAMadd#1{#1}
\def\CAMdel#1{}

\begin{document}
\title{Xavier: Toward Better Coding Assistance in Authoring Tabular Data Wrangling Scripts}

\author{Yunfan Zhou}
\orcid{0009-0009-7814-4390}
\affiliation{%
    \institution{State Key Lab of CAD\&CG,\\Zhejiang University}
    \city{Hangzhou}
    \state{Zhejiang}
    \country{China}
}
\email{yunfzhou@zju.edu.cn}

\author{Xiwen Cai}
\orcid{0000-0002-7256-4660}
\affiliation{%
    \institution{State Key Lab of CAD\&CG,\\Zhejiang University}
    \city{Hangzhou}
    \state{Zhejiang}
    \country{China}
}
\email{xwcai@zju.edu.cn}

\author{Qiming Shi}
\orcid{0009-0002-7876-1056}
\affiliation{%
    \institution{State Key Lab of CAD\&CG,\\Zhejiang University}
    \city{Hangzhou}
    \state{Zhejiang}
    \country{China}
}
\email{qimingshi@zju.edu.cn}

\author{Yanwei Huang}
\orcid{0009-0001-9453-7815}
\affiliation{%
    \institution{State Key Lab of CAD\&CG,\\Zhejiang University}
    \city{Hangzhou}
    \state{Zhejiang}
    \country{China}
}
\email{huangyw@zju.edu.cn}

\author{Haotian Li}
\orcid{0000-0001-9547-3449}
\authornote{The work was done when Haotian Li was at HKUST.}
\affiliation{%
    \institution{Microsoft Research Asia}
    \city{Beijing}
    \country{China}
}
\email{haotian.li@microsoft.com}

\author{Huamin Qu}
\orcid{0000-0002-3344-9694}
\affiliation{%
    \institution{The Hong Kong University of Science and Technology}
    \city{Hong Kong SAR}
    \country{China}
}
\email{huamin@cse.ust.hk}

\author{Di Weng}
\orcid{0000-0003-2712-7274}
\authornote{Di Weng is the corresponding author.}
\affiliation{%
    \institution{School of Software Technology,\\Zhejiang University}
    \city{Ningbo}
    \state{Zhejiang}
    \country{China}
}
\email{dweng@zju.edu.cn}

\author{Yingcai Wu}
\orcid{0000-0002-1119-3237}
\affiliation{%
    \institution{State Key Lab of CAD\&CG,\\Zhejiang University}
    \city{Hangzhou}
    \state{Zhejiang}
    \country{China}
}
\email{ycwu@zju.edu.cn}

\renewcommand{\shortauthors}{Zhou et al.}

\begin{abstract}

Data analysts frequently employ code completion tools in writing custom scripts to tackle complex tabular data wrangling tasks.
\DIFdel{However, existing tools typically provide completions based solely on code, overlooking the importance of data contexts such as schemas and values.}
\DIFadd{However, existing tools do not sufficiently link the data contexts such as schemas and values with the code being edited.}
This not only leads to poor code suggestions, but also frequent interruptions in coding processes as users need additional code to locate and understand relevant data.
We introduce \textit{Xavier}, a tool designed to enhance data wrangling script authoring in computational notebooks.
\textit{Xavier} maintains users' awareness of data contexts while providing data-aware code suggestions.
It automatically highlights the most relevant data based on the user's code, integrates both code and data contexts for more accurate suggestions, and instantly previews data transformation results for easy verification.
To evaluate the effectiveness and usability of \textit{Xavier}, we conducted a user study with 16 data analysts, showing its potential to streamline data wrangling scripts authoring.

\end{abstract}

\ccsdesc[500]{Human-centered computing~User interface programming}

\keywords{Interactive data wrangling, coding assistance}

\begin{teaserfigure}
    \centering
    \includegraphics[width=\linewidth]{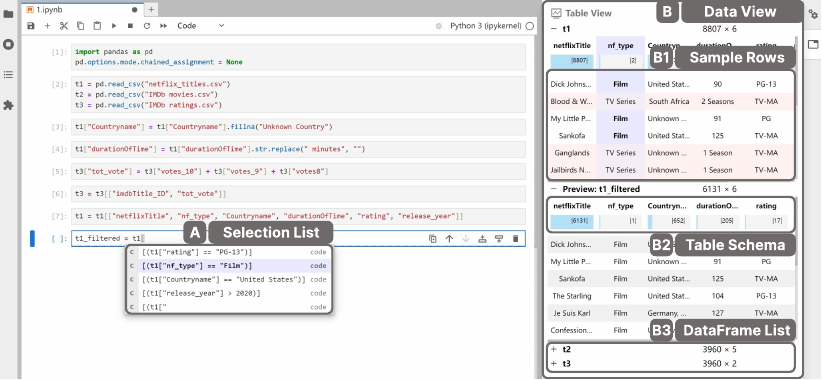}
    \caption{The interface of \textit{Xavier} within the computational notebook. A) The selection list providing data context-aware code completions for users while typing. B) The data view displaying data contexts for user reference. During typing, \textit{Xavier} automatically expands items by showing the table schema (B2) and sample rows (B1), highlighting the most relevant data based on user's code. Other irrelevant data contexts are listed in the data view and folded (B3).}
    \Description{A computational notebook interface with the integration of Xavier to illustrate the overall interface design. A) A floating selection list appears under the user's code with several code completion suggestions as users type. B) The data view in the side panel of the notebook lists DataFrames active in the notebook memory. The item can be expanded. Table schema and sample rows are displayed in a spreadsheet style.}
    \label{fig:ui-overview}
\end{teaserfigure}

\maketitle

\section{Introduction}

Data wrangling is a crucial process in data science that involves tasks such as data cleaning, integration and format transformation for downstream analysis~\cite{wrangler}.
To perform complex tabular data wrangling tasks, data analysts with programming skills often write ad-hoc scripts.
Given the tedious and time-consuming nature of scripting, the emergence of various code completion tools, particularly AI-powered ones including IntelliSense~\cite{intellisense} and GitHub Copilot~\cite{github-copilot}, has significantly enhanced coding efficiency in programming environments.
\DIFdel{Despite their widespread application, code completion tools have a fundamental limitation in their narrow focus on code-related features (e.g., grammar and semantics), or \textit{code contexts}, in the completion process.
The metadata of the datasets people are working with, such as table schemas and unique values of columns, is essential in data wrangling scripting, yet it is often overlooked by the existing code completion tools.
Our preliminary study with expert data analysts reveals that without awareness of such \textit{data contexts} while writing the code, users may face difficulties in locating relevant data for wrangling tasks, ending up frequently searching for data tables, filtering unnecessary data pieces, and printing sample data in notebooks.
With mere access to code text rather than the underlying data context, these completion tools fail to generate recommendations that require a strong understanding of datasets, such as correct column names to join two tables, or the prefix of a data value to be removed.}

\DIFadd{Despite their widespread application, AI-powered code completion tools primarily focus on code-related features (e.g., grammar and semantics), or \textit{code contexts}, in the completion process.
These tools often overlook the metadata of datasets people are working with, or \textit{data contexts}, leading to issues when generating code.
For instance, since the underlying Large Language Models (LLMs) may memorize training examples~\cite{extract-training-data}, AI-powered code completion tools often generate wrong columns or unreasonable values~\cite{waitgpt}.
Some AI-powered code completion tools, like GitHub Copilot which has access to original data files in the project~\cite{copilot-understanding}, consider datasets to a certain extent.
However, they do not sufficiently link the metadata (e.g. table schemas and unique values of columns) to the specific code being edited.
Therefore, these completion tools often fail to generate recommendations that require a strong understanding of datasets, such as correct column names to join two tables, or the prefix of a data value to be removed.}

\DIFdel{The limitations of existing tools motivate us to propose a novel coding assistance approach that prioritizes data contexts, allowing users to remain continuously aware of data contexts while providing data context-aware code suggestions.
However, developing such an approach presents several challenges.
First, incorporating data context for intelligent code completion remains unexplored, since code completion for different data wrangling operators requires different types of data contexts.
Additionally, locating specific data segments that are relevant and useful for users is challenging due to the dynamic nature of user focus during coding.
Moreover, users face challenges in validating the accuracy of AI-driven code completions according to prior work~\mbox{\cite{grounded-copilot}} and our preliminary study, indicating that a simplified process for code verification is desired.
}

\DIFadd{The limitations of existing tools motivate us to propose a novel coding assistance approach that links data contexts to the code being edited.
However, developing such an approach presents several challenges according to our preliminary study.
First, users are often frustrated at mistakes of AI-generated code and have to manually specify dataset information for intelligent code completion.
However, dynamically linking data contexts to reduce the effort of data context specification remains unexplored.
Additionally, locating specific data segments that are relevant and useful for users is challenging due to the dynamic nature of user focus during editing the incomplete code.
Moreover, users have to spend considerable time validating the transformation result of AI-generated code~\cite{autoprofiler, lp-validate-ai-cc}.
Existing live programming tools like Projection Boxes~\cite{projection-boxes} can display information of variables (e.g. runtime values) in a real-time way as code is being edited, helping users validate the code to some extent.
However, they do not link the most relevant information in the DataFrame variable to the code being edited, since users often focus on a small portion of information in the variable while editing data wrangling code.
}
To tackle these challenges, we use Pandas~\cite{pandas} as an exemplary data transformation library, delineating a strategy combining code contexts and data contexts to provide data context-aware code completions.
Additionally, we propose a method to dynamically link code and data, highlight the relevant data based on user's code, and allow instant preview of data transformation results for straightforward verification.
Finally, an user interface is designed, leveraging visualizations and interactions to facilitate user's data sensemaking in the scripting process.
On top of this, we take the computational notebook as an exemplary programming environment that facilitates trying and iterating data wrangling code~\cite{exploration-explanation-notebooks} and introduce \textit{Xavier}, a coding assistance extension for computational notebooks.
To evaluate \textit{Xavier}'s usability and effectiveness, we conducted a counterbalanced mixed-design user study on 16 participants, in which participants wrote data wrangling scripts to complete a task with \textit{Xavier} and another task with a baseline tool.
We find that users experienced significantly fewer context switches and errors during scripting by using \textit{Xavier}.
User feedback further confirms that \textit{Xavier} helps users maintain data context awareness and author code more smoothly throughout the data wrangling scripting process.
The major contributions are summarized as follows:

\begin{itemize}
    \item A preliminary study that reveals user behavior patterns in authoring tabular data wrangling scripts and summarizes user requirements of \DIFadd{AI-powered} coding assistants.
    \item A computational notebook extension, \textit{Xavier}, which aids users in maintaining awareness of data \DIFdel{context}\DIFadd{contexts} during data wrangling code authoring.
\end{itemize}

\section{Related Work}
In this section, we review research on data wrangling tools, coding assistants in computational notebooks\DIFdel{and}, code completion tools \DIFadd{and live programming tools.}

\subsection{Data Wrangling Tools}

Data wrangling is a challenging task that involves change, rearrangement, and merging of data to prepare for various purposes such as visualization and analysis~\cite{wrangler}.
Several toolkits like Pandas~\cite{pandas} in Python, as well as dplyr~\cite{dplyr} and tidyr~\cite{tidyr} in R, have been developed to facilitate this process.
These toolkits offer data analysts with significant flexibility for effective data manipulation.
Nevertheless, analysts unfamiliar with Python or R may find it time-consuming and demanding to learn a new toolkit.
To address these challenges, numerous interactive tools have been proposed to lower the barriers associated with data wrangling.

Interactive tools for tabular data wrangling can be categorized as either \textit{imperative} or \textit{declarative}.
\textit{Imperative} approaches~\cite{potter, wrangler, proactive-wrangler} focus on the wrangling procedures and typically provide users with a menu of various wrangling operations.
In contrast, \textit{declarative} approaches emphasize the specification of transformation outcomes.
For instance, many systems allow users to compose example target tables and automatically synthesize transformation programs~\cite{foofah, flashfill2, falx}.
Other tools enable users to specify the intended tasks through declarative mappings~\cite{rigel} or natural language~\DIFdel{\mbox{\cite{nl2rigel}}}\DIFadd{\cite{nl2rigel, coldeco}}.
While these interactive tools help democratize the wrangling process, they are primarily designed for users with limited experience in data wrangling and programming.
Writing custom wrangling scripts remains common among data analysts, as coding offers a more flexible and expressive way to specify the wrangling process~\cite{how-data-science-workers, how-data-worker-collaborate}.
Hence, \textit{Xavier} offers coding support for data analysts proficient in programming.

\DIFdel{
Additionally, an extensive range of earlier studies focus on code recommendation for data wrangling.
Tools like CoWrangler~\mbox{\cite{cowrangler}} and SnipSuggest~\mbox{\cite{snipsuggest}} propose heuristic methods to generate recommended code, while tools like AutoPandas~\mbox{\cite{autopandas}} and Auto-Suggest~\mbox{\cite{auto-suggest}} leverage deep neural networks to learn patterns of transformation operator usage and provide more intelligent code recommendations.
However, most of these tools focus on recommendation algorithms or models and are standalone, requiring an extra panel to input data and specify transformations before exporting scripts.
Since data wrangling is an iterative process~\mbox{\cite{wrangling-survey}}, programmers often need to work between the extra panel and their own code editor (e.g. copy and paste snippets, import and export data), leading to considerable context-switch overhead.
In contrast, \textit{Xavier} offers coding assistance within computational notebooks, supporting awareness of data context to expedite code editing.
}
\DIFadd{
Additionally, earlier studies have focused on code recommendation for data wrangling, which can be categorized into three types according to the smallest granularity of the output code: \textit{block-level}, \textit{line-level} and \textit{token-level}.
\textit{Block-level} code recommendation tools~\cite{cowrangler, autopandas, step-wise-phase-wise} generate blocks of data wrangling code that often include multiple transformation operations.
For instance, AutoPandas~\cite{autopandas} leverages program synthesis techniques to generate code from input-output examples.
Stepwise~\cite{step-wise-phase-wise} and Phasewise~\cite{step-wise-phase-wise} decompose the entire data wrangling problem into steps or phases to facilitate steering and verification of AI-generated code. 
\textit{Line-level} code recommendation tools like Auto-Suggest~\cite{auto-suggest} focus on generating single-operation data wrangling code.
\textit{Token-level} code recommendation tools like SnipSuggest~\cite{snipsuggest} support suggesting next few tokens such as table names or column names as users write part of the code.}

\DIFadd{
Although \textit{token-level} code recommendation tools basically incorporate data contexts into their recommendation models, users still face challenges in understanding and verifying the recommendations.
In contrast, \textit{Xavier} not only links relevant data contexts to the code being edited in its recommendation model, but also offers corresponding highlighting and previews to provide on-the-fly explanations for its recommendations.
}

\subsection{Code Assistants in Computational Notebooks}

\DIFadd{
A vast range of code assistants have been proposed for integration into computational notebooks, aiming to enhance the productivity of data workers in programming tasks.
These assistants can be categorized based on stages of data analysis in which they provide recommendations: \textit{exploration} (i.e. during the data exploration stage), \textit{next-step} (i.e. suggesting the next steps before typing), or \textit{typing} (i.e. offering assistance during typing).
}

\DIFadd{
\textit{Exploration} code assistants~\cite{edassistant, lodestar} help data workers discover workflows or analysis techniques during data exploration.
For example, EDAssistant~\cite{edassistant} aids users by finding code snippets similar to those in the current notebook through in-situ code search, thereby inspiring the workflows in their analysis.
\textit{Next-step} code assistants~\cite{mage, wrex, jigsaw, data-wrangler, biscuit} specialize in recommending next steps before users start typing, guiding users to continue their workflows effectively.
For instance, Wrex~\cite{wrex} provides spreadsheet-like interfaces for data manipulation by example.
Jigsaw~\cite{jigsaw} leverages large language models to enable multi-modal inputs for data wrangling script generation.
BISCUIT~\cite{biscuit} scaffolds users understanding and refining the LLM-generated code by introducing ephemeral UIs.
\textit{Typing} code assistants~\cite{glinda, github-copilot} offer real-time recommendations while users are typing.
For example, Glinda~\cite{glinda} combines live programming, GUI elements, and a Domain-Specific Language (DSL) to provide immediate feedback during programming.
}

\DIFadd{However, few code assistants focus on real-time recommendations during \textit{typing}.
Although Glinda~\cite{glinda} incorporates a code completion feature, it is primarily designed for the rapid construction of DSL structures based on language grammar.
GitHub Copilot~\cite{github-copilot} does not sufficiently link the data contexts to the specific code being edited, which will be discussed in Section~\ref{ssec:code-completion-tools}.
In contrast, \textit{Xavier} focuses on code completions for data wrangling scripts, allowing users to remain aware of data contexts during typing.
}

\subsection{Code Completion Tools}
\label{ssec:code-completion-tools}

Code completion, which suggests candidate subsequent tokens for programmers, is a widely used feature in code editors that accelerates the programming process.
It has been extensively investigated in the literature~\cite{code-comp-rw}.
Early code completion approaches relied on heuristic rules, such as static type information~\cite{bcc3, type-partial-exp, insynth}, similar patterns from codebases~\cite{learn-example-ccs, auto-method-comp, grapacc}, or usage frequency~\cite{program-history-complete, program-history-complete2, multi-version-comp}.
Building on the analogy between human-written code and natural language~\cite{naturalness}, subsequent works~\cite{slang, slp-core, deep3} explored statistical approaches that leveraged the repetitive and predictive properties of code.
With the advancement of deep neural networks, language models based on neural networks, such as Recurrent Neural Networks (RNNs) and Long Short-Term Memory (LSTM), have also been investigated~\cite{pointer-network, pythia, ovm-comp} to improve the suggestion accuracy.

More recently, as Generative Pre-trained Transformers (GPT) gained popularity in natural language processing, numerous GPT-based models and tools have emerged~\cite{cuglm, codex, intellicode-compose}.
For example, fine-tuned on a large corpus of publicly available code on GitHub, Codex~\cite{codex} demonstrated improved accuracy in code completion.
Its production version, GitHub Copilot~\cite{github-copilot}, is capable of providing longer code completions, ranging from individual tokens to entire functions, while maintaining a relatively high level of accuracy.

These approaches optimize the exploration and utilization of deeper semantic information in code, primarily supporting multilingual code completion and general tasks.
Nevertheless, data contexts such as schemas and values are also crucial for enhancing accuracy in data wrangling code authoring.
Besides, most code completion tools focus on coding assistance during code authoring, which can cause users to lose data contexts before or after code authoring, making it difficult to locate relevant data and verify code.
Therefore, \textit{Xavier} incorporates both code contexts and data contexts for more intelligent code completions, and provides assistance for awareness of data contexts throughout the coding process.

\subsection{Live Programming Tools}

\DIFadd{Live programming provides real-time feedback (e.g. visualization of a program's runtime data) as code is edited~\cite{lp-validate-ai-cc}, which enhances understanding of how code changes affect the running system~\cite{omnicode}. 
General tools like Projection Boxes~\cite{projection-boxes} and Engraft~\cite{engraft} offer configurable frameworks for live programming.
Other tools focus on specific programming tasks and can be categorized into three types: \textit{generation}, \textit{debugging}, and \textit{validation}.}

\DIFadd{\textit{Generation} tools~\cite{small-step-lp, loopy} allow users to modify runtime values to synthesize code.
For instance, SnipPy~\cite{small-step-lp} supports small-step program synthesis by changing the displayed runtime values.
\textit{Debugging} tools~\cite{omnicode, oden} aid in program debugging by visualizing runtime values.
For example, OmniCode~\cite{omnicode} displays the entire history of all run-time values for all program variables all the time when users are writing Python code.
\textit{Validation} tools~\cite{lp-validate-ai-cc, ivie} enhance validation of AI-generated code by providing continuous explanations adjacent to the code.
For instance, Ivie~\cite{ivie} breaks up complex code into pieces and annotates them with textual explanations.
These prior works have made efforts to improve code understanding in programming, inspiring the design of \textit{Xavier}.}
\DIFadd{However, when users are editing code to apply a new transformation on datasets, they typically focus on a relatively small portion of information in variables.
For instance, while applying a string format transformation like \mycode{\lstinline{@df@["A"] \ = @df@["A"].~str~.replace(":", "")}} on the variable ``df'' with many columns, users typically focus merely on the value format of the column ``A'' in the variable ``df''. 
Although existing \textit{validation} tools can effectively display the necessary information of variable states (e.g. all elements in an array), they generally fail to link such a small portion of relevant information required to validate the current code that users are authoring.
Our work offers a nuanced inspection by highlighting the relevant data contexts and showing instant previews, which further enhances live programming and code understanding.}

\section{Preliminary Study}
\DIFadd{%
Before designing an AI-powered code assistant for data wrangling scripts, we would like to understand the requirements and challenges the users may meet.
Prior work~\cite{grounded-copilot, reading-lines, expect-experience, large-scale-survey-copilot, program-with-ai} has explored patterns of user interactions with AI-powered code assistants in general coding tasks and identified various challenges, such as code understanding, verification, and context switching.
These efforts have pointed out the direction for improving AI-powered code assistants and have inspired our research.
}
\DIFadd{
Although the findings from the existing work can be generalized to various domains, they do not fully cover the problems users encounter when authoring data wrangling scripts.
Unlike tasks with clear objectives such as front-end and back-end development or web scraping, data wrangling tasks often involve a higher degree of uncertainty. 
Users need to fully explore and understand the characteristics of the dataset~\cite{ferry} and complete data transformation tasks iteratively through interacting with data~\cite{autoprofiler}. 
In order to understand the needs and challenges users face when using AI-powered code assistants to complete data wrangling tasks,}
\DIFdel{To better design a code assistant tool for data wrangling scripts,}we conducted a preliminary study\footnote{The study has been approved by State Key Lab of CAD\&CG, Zhejiang University.} \DIFdel{to understand users' behavior and needs when authoring data wrangling scripts.
The}\DIFadd{where} participants took part in: 1) an experiment, in which they were asked to complete data wrangling tasks using GitHub Copilot~\cite{github-copilot}, and
2) a semi-structured interview, in which they shared the basic workflow of data wrangling and the issues encountered in the previous experiment.
We identified common patterns in users' behavior and summarized requirements according to the interview.

\subsection{Participants}

We recruited 9 data analysts (denoted as P1-P9, 5 male and 4 female, $Age_{mean}=26.44$, $Age_{std}=5.88$, $Experience_{mean}=5.00\ years$, $Experience_{std}=2.55\ years$) by sending invitations via social media and word-of-mouth.
They had diverse backgrounds such as Control Engineering, Mathematics, Digital Media Technology, Data Governance, Geographic Information Systems and so on.
They regularly programmed with Python Pandas in computational notebooks for data wrangling in their projects (at least once a week).
Their demographic information is detailed in supplemental materials.
Participants consented to having their voices and programming processes recorded.

\subsection{Apparatus and Materials}
\label{ssec:preliminary-apparatus-material}

\hspace*{\parindent}\textbf{Apparatus.}
We chose GitHub Copilot~\cite{github-copilot}, a representative of AI-assisted code assistant, in our preliminary study.
Copilot is the first code assistant based on Large Language Models (LLMs) to reach widespread usage and has been studied by a wide range of literature~\cite{grounded-copilot, design-code-assistant, lp-validate-ai-cc}.
Participants authored data wrangling code in computational notebooks on standardized desktop devices.
A slide with task descriptions, output examples and data dictionaries was prepared for users to refer to in the code authoring experiment.

\textbf{Datasets.}
We selected \textit{Covid-19} dataset including 3 tables and 20 columns based on a publicly available notebook\footnote{\url{https://www.kaggle.com/code/erikbruin/storytelling-covid-19}}.
The dataset was chosen for its public familiarity~\cite{task-based-effect-vis} and coverage of two major types of tabular data, namely categorical and numerical.
Data tables were slightly modified such as changes of column names and categorical values to ensure Copilot had not been trained on them.

\textbf{Tasks.}
In our preliminary study, participants were asked to author a data wrangling script to calculate the ratio of confirmed cases and death cases for each country.
They needed to complete approximately 20 transformation operations, involving common operations such as joining, sorting, and filtering, to produce an output table that includes 7 columns.
Details of task description are left to the supplementary materials.

\subsection{Procedure}
\label{ssec:pre-study-procedure}

We first informed participants about relevant information regarding the study, including the purpose, overall procedure, and compensation of the study, and then sought participants' consent to data collection.
Then, the participants would take part in a code authoring experiment (60-75 minutes) and a semi-structured interview (15 minutes).
We captured participants' behavior in coding process through video recordings and collected feedback in the interview by audio recordings for subsequent analysis and summarization.
The entire study took around 90 minutes and each participant received 70 Chinese Yuan as compensation.

\textbf{Code authoring experiment.}
Initially, we briefly introduced GitHub Copilot and allowed participants to try it on a warm-up task, in order to ensure participants grasped its usage.
Participants were then given the task description, dataset and data dictionary to write a script in a computational notebook with Copilot's assistance.
Participants were free to consult the data dictionary or view raw data tables as needed.
Following the think-aloud protocol, participants were encouraged to share their thoughts on Copilot during the experiment.

\textbf{Semi-structured interview.}
The semi-structured interview consisted of three parts.
In the initial part, we inquired of participants regarding their basic workflow of data wrangling in their data analysis work.
In the second part, participants were encouraged to share the issues they encountered in the experiment, including the difficulties in completing the task and the pain points of interacting with Copilot.
In the last part, we specifically asked participants about the issues we observed in the experiment.

\subsection{Findings}
\label{sec:study-findings}

\begin{figure*}[t]
  \centering
  \includegraphics[width=0.95\linewidth]{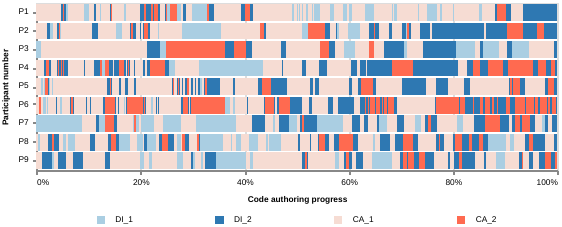}
  \caption[]{
    The timelines of observed activities \DIFadd{of each participant} during \DIFadd{the} code authoring experiment. \textbf{DI} and \textbf{CA} refer to \textbf{Data Inspection} and \textbf{Code Authoring}, respectively. Users may profile data (\dione{DI\_1}), verify results (\ditwo{DI\_2}), create new transformations (\caone{CA\_1}) or modify the written code (\catwo{CA\_2}) during scripting. \DIFadd{To facilitate comparison, we scaled the duration of activities by normalizing each participant's total time spent on code authoring.}
  }
  \Description{A gantt chart to illustrate changes of observed activities during code authoring experiment in our preliminary study. Different types of activities are encoded using different colors: DI_1 (light blue), DI_2 (dark blue), CA_1 (light red), CA_2 (dark red).}
  \label{fig:codingProcess}
  \addvspace{-7pt}
\end{figure*}

\DIFadd{We manually captured participants' behaviors in the code authoring experiment through video recordings.
To analyze the user feedback from the semi-structured interview, \CAMadd{we conducted a qualitative inductive content analysis~\cite{content-analysis}.}
\CAMadd{Initially, }two co-authors (data analysis experience: $\geq 3\ years$) read the recording transcripts to extract the comments related to issues and difficulties individually.
Then, they grouped the similar comments and identified the challenges collaboratively.
Finally, all co-authors discussed the user behavior and challenges to derive the key findings.}
\DIFdel{In this subsection, we first introduce the action types identified from participants' behavior and then introduce the findings according to our observations and the participants' feedback.}
\DIFadd{In this subsection, we first introduce the definition of two kinds of participants' activities observed in the code authoring experiment.
Then we summarize four findings according to user activities and feedback.}

\subsubsection{Definition of Two Kinds of \DIFdel{Actions}\DIFadd{Activities}}
\label{sssec:def-di-ca}
We categorized the participants' behavior into two kinds of \DIFdel{actions}\DIFadd{activities}: \textbf{Data Inspection (DI)} and \textbf{Code Authoring (CA)}.
\textbf{DI} involves examining datasets, which can be further divided into two types: profiling (\dione{DI\_1}) and verifying (\ditwo{DI\_2}).
\dione{DI\_1} is focused on understanding datasets and gaining inspirations for subsequent wrangling steps, which can manifest in multiple ways such as reading parts of the data table.
\ditwo{DI\_2} is about verifying the effectiveness and correctness of transformations, typically manifested as viewing the result table after running the transformation code.
\textbf{CA} involves writing code, which can also be divided into creating (\caone{CA\_1}) and modifying (\catwo{CA\_2}).
\caone{CA\_1} is related to applying new transformations or profiling datasets, while \catwo{CA\_2} involves minor code modifications, such as adjusting function parameters, often seen in debugging.
\DIFadd{We measured duration of activities for each participant and scaled the duration of activities by normalizing each participant’s total time spent on code authoring.}
\DIFdel{The activity timeline of each participants during code authoring experiment is shown in \mbox{\autoref{fig:codingProcess}}.}
\DIFadd{The activity timelines of each participant during code authoring experiment are shown in \autoref{fig:codingProcess}.}

\subsubsection{User Behavior and Challenges}
\textcolor{black}{}
\DIFdel{We summarize common patterns of user behavior from the observation of our code authoring experiment and challenges according to the semi-structured interview.}
\DIFadd{
According to the user behavior and feedback, we summarized four findings:
}

\textit{Users frequently switched between code contexts and data contexts, commonly fixing coding mistakes by checking data.}
According to \autoref{fig:codingProcess} and our observations in the code authoring experiment, participants frequently switched between \dione{DI\_1} and \caone{CA\_1} or between \ditwo{DI\_2} and \caone{CA\_1}, indicating frequent context switches during the coding process.
For instance, P1 encountered a relatively difficult subtask in the second half of code authoring experiment, and he needed to write code (\caone{CA\_1}) and review the data (\dione{DI\_1}) iteratively to adjust his coding approach.
Ideally, users who author data wrangling scripts experience an iterative process involving profiling, creating, and verifying steps, which typically follow the cycle: \dione{DI\_1}, \caone{CA\_1}, \ditwo{DI\_2}, (\dione{DI\_1}), \caone{CA\_1}, \ditwo{DI\_2},...
However, the interleaving of \catwo{CA\_2} and \ditwo{DI\_2} is common in \autoref{fig:codingProcess}, indicating that participants often encountered errors and typically fixed mistakes by checking the result table.
As a representative example, P6 encountered an intractable bug in her script in the last third of the experiment, having to repeatedly modify the code (\catwo{CA\_2}) and examine the result (\ditwo{DI\_2}).

\textit{Users complained about frequent mistakes brought by Copilot's code completion.}
When writing partial code and letting Copilot complete it, doubts arose about its data knowledge.
In Section \ref{ssec:preliminary-apparatus-material}, all column names were changed (e.g. ``ProvinceOrState'' instead of ``Province/State''), yet Copilot often returned incorrect completions like ``Province/State''.
Such mistakes caused ``non-exist column'' errors which frustrated participants.
``\textit{Why did Copilot still complete `Province/State' even if I explicitly mentioned the `ProvinceOrState' column in previous code?}'', P8 questioned.
When preferring to write comments first for Copilot to generate code, participants complained about lengthy prompts.
P2, P4 and P9 admitted their detailed dataset information in comments was mainly for Copilot, which echoed previous research findings~\cite{grounded-copilot}.
P2 doubted Copilot would generate correct code without specifying operations and operators.
Nearly all participants suggested Copilot should access data tables so as to correctly complete column names and values.
\textit{Users faced difficulties searching for the part of the data to which the code completion tool was referring.}
Before applying a new transformation operation, Copilot recommended the next possible line of code for participants. 
However, participants usually had no idea about the part of the data Copilot was focusing on and they were not sure about the relevance between the next-step recommendation and their current wrangling subtasks.
Hence, they tended to reject the recommendation and instead used various methods to search datasets.
For example, before unifying country names, participants wrote additional code to filter relevant names (P1, P4 and P6) or opened original CSV files (P2, P3, P5, P7 and P8) to search globally (\dione{DI\_1}).
However, they found it time-consuming to search, as P2 commented, ``\textit{I was overwhelming when searching for `United States' in the raw data file with so much irrelevant and interfering information}''.
P5 noted, ``\textit{It was troublesome writing code just to see different spellings of United States, such as `US' and `USA'}.''
Interruptions in writing often occurred to observe intermediate variables during transformations.
For example, P1 printed columns of tables \mycode{\lstinline{@covid_19_data@}} and \mycode{\lstinline{@country_codes@}} to recall the exact spelling of key names while merging the two tables.

\textit{Users spent considerable time reviewing the transformation result of AI-generated code.}
Due to frequent ``non-exist column'' errors, participants remained skeptical especially about constants like column names or data values, requiring careful verification against data.
However, reviewing was sometimes inconvenient even when running code.
For instance, P6 spent about 10 minutes fixing a line due to unawareness of a column's data type and a hidden spelling error.
``\textit{I expected cardinality changes of the column after the string replacing transformation, but needed extra code to verify the prediction.}''
P1's comment indicated the challenge in tracking data changes, as reported in~\cite{ditl, autoprofiler}.

\subsection{User Requirements}
\label{ssec:user-requirements}

We identified three requirements for a code completion tool in authoring data wrangling scripts according to findings.
In the remaining part of the paper, \textbf{R1-R3} refers to the requirements.
The requirements are:

\textbf{R1. Incorporating data contexts for intelligent code completion.}
In our code authoring experiment, users were frustrated at mistakes (e.g. non-exist columns) made by Copilot and had to write lengthy prompts to transfer the dataset information to Copilot.
However, general code completion tools like Copilot typically lack data contexts, which limits the intelligence of suggestions.
Thus, automatic data context provision can help alleviate the need to manually convey data contexts to the code completion tool.

\textbf{R2. Assisting users in locating relevant parts of datasets.}
In our code authoring experiment, participants struggled to search for relevant data to be wrangled and they had to frequently print tables or refer to original CSV files, which caused a significant context-switch overhead.
Therefore, the tool should offer a straightforward way to help users search for and focus on relevant data before they start writing code to apply a new transformation.

\textbf{R3. Offering straightforward code verification.}
In our code authoring experiment, participants were confused by \DIFdel{the recommendation principle of Copilot}\DIFadd{how Copilot suggested code completion} and took much effort in verifying correctness of Copilot's suggestions, which also lowered the efficiency of code authoring.
Thus, the tool should also offer simple verification approaches to assist users in \DIFdel{understanding the recommendation principle}\DIFadd{validating AI-generated code} and enhance \DIFdel{users' trust in the suggested code}\DIFadd{their trust}.

\section{\textit{Xavier}}
\label{sec:xavier}

In this section, we first present an overview of \textit{Xavier} (Section~\ref{ssec:overview}).
This is followed by a detailed discussion of the design of \textit{Xavier}, which features data context-aware code completion (Section~\ref{ssec:code-comp}), automatic data context highlighting (Section~\ref{ssec:data-ctx-highlight}), and real-time transformation preview (Section~\ref{ssec:data-cxt-preview}).
\DIFadd{Usage scenarios in Section~\ref{ssec:code-comp}, \ref{ssec:data-ctx-highlight} and \ref{ssec:data-cxt-preview} are highlighted with a \colorbox{scenariocolor}{blue background}.}

\subsection{Overview}
\label{ssec:overview}

\begin{figure*}[t]
  \centering
  \includegraphics[width=\linewidth]{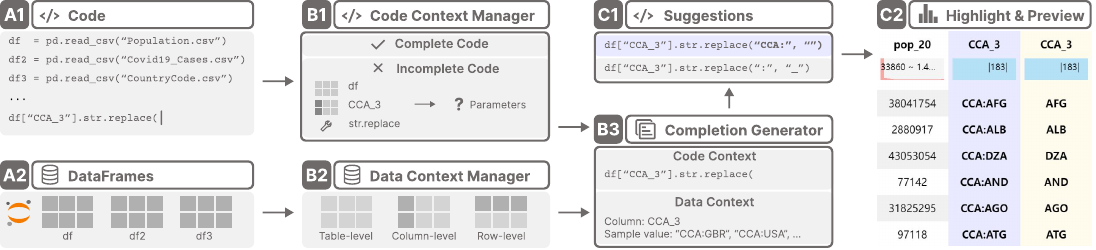}
  \caption{
    The workflow of \textit{Xavier}. The input of \textit{Xavier} consists of code in the editor (A1) and DataFrames in the notebook kernel (A2). The code is divided into the complete part and the incomplete part by the code context manager (B1) where the incomplete code is further parsed. Data contexts for each DataFrame are pre-calculated in the data context manager (B2) since the last run of code. The complete code, the parsing result and data contexts are transferred to completion generator (B3) for data context-aware code suggestions (C1). Meanwhile, \textit{Xavier} highlights the most relevant data based on user's code and the completion suggestions in the data view, previewing transformation results to facilitate code verification (C2).
  }
  \Description{Seven components are connected in a style of dataflow diagram to illustrate the workflow of Xavier. The flow chart starts with Code and Data in the notebook and ends with completion suggestions, highlight and preview in the user interface.}
  \label{fig:workflow}
  \addvspace{-7pt}
\end{figure*}

\textit{Xavier} is a coding assistance tool enabling users to stay\CAMdel{continuously} aware of data contexts while authoring data wrangling scripts.
Integrated as an extension for computational notebooks, it supports Python Pandas~\cite{pandas}, with the potential for generalization to other data transformation libraries.
\CAMdel{Data contexts are integrated into the design of \textit{Xavier}.
Inspired by a previous study~\mbox{\cite{autoprofiler}}, we categorize data contexts into three types according to three different types of data object~\mbox{\cite{table-scraps}}: tables, rows and columns.
Table-level data contexts contain \mbox{\DIFdel{metadata}\DIFadd{basic information}} of a data table, including table name, column name and shape.
Column-level data contexts include detailed information of a column.
Common column-level data contexts are data type, null value count and sortedness.
In addition to common column-level contexts, categorical columns have unique values, value frequency, cardinality and value format, while numerical columns have value range, sample data points and value format.
Row-level data contexts include sample rows of a table, maintaining the spatial relationship between table cells.
Different data contexts can be combined and enhance functionalities in \textit{Xavier} like code completion.
For instance, column names from table-level contexts and sample rows from row-level contexts are both required for the table join operation \mbox{\DIFdel{for completion of}\DIFadd{to complete}} parameters indicating the key columns.
The same \mbox{\DIFdel{level}\DIFadd{type}} of data contexts of different data objects can also be combined.
For example, table-level data contexts from multiple \mbox{\DIFdel{different}} tables are essential for the table concatenation operation.}

\autoref{fig:ui-overview} illustrates \textit{Xavier} within the notebook interface.
It consists of two components: the notebook interface and an always-on data view visualizing data contexts.
In parallel to users' writing scripts in the notebook, a selection list (\autoref{fig:ui-overview} A) will appear, providing completion suggestions for users.
Meanwhile, the data view (\autoref{fig:ui-overview} B) displays data contexts for user reference.
\CAMdel{Since the visualization of data contexts (e.g. multi-table data contexts for the \mbox{\mycode{\lstinline{@df@.merge}}} operation) may occupy a significant amount of screen space, the data view is kept in a split window on the right. 
This layout choice minimizes visual obstruction of the code editor and reduces repeatedly scrolling to view the code.}
Initially, all active DataFrames in the notebook kernel are listed (\autoref{fig:ui-overview} B3).
Expanding an item in the DataFrame list reveals the schema (\autoref{fig:ui-overview} B2) with profiles such as data type, cardinality and value range of each column, helping users rapidly recall the content of the DataFrame.
Clicking the ``Show sample rows...'' button displays the first 15 rows \DIFdel{of the DataFrame}(\autoref{fig:ui-overview} B1), aiding further understanding without repeatedly using additional code like \mycode{\lstinline{@df@.head()}} to print DataFrames.
It also facilitates efficient verification of analysis results by automatically highlighting \CAMadd{and previewing} real-time data contexts based on the current partial code in the editor and the selected item in the completion list.
\CAMadd{Since the visualization of data contexts may occupy a significant amount of screen space, the data view is kept in a split window on the right. 
This layout choice minimizes visual obstruction of the code editor and reduces repeatedly scrolling to view the code.}
These components cover two major kinds of activities (i.e. \textbf{DI} and \textbf{CA}) discussed in Section \ref{sec:study-findings}, keeping users aware of data contexts throughout the coding process.
To support the functionalities of the interface, three computational components, namely \textit{code context manager}, \textit{data context manager} and \textit{completion generator} are designed.
Their relationship is shown in \autoref{fig:workflow}.
The \textit{code context manager} (\autoref{fig:workflow} B1) splits the code in the current notebook (\autoref{fig:workflow} A1) into the complete part and incomplete part, parsing the partial code near the input cursor.
The parsing result serves as clues to gather relevant data contexts for the subsequent workflow.
The \textit{data context manager} (\autoref{fig:workflow} B2) computes data contexts for all active DataFrames in the notebook kernel (\autoref{fig:workflow} A2).
Both code and data contexts are then transferred to the \textit{completion generator} (\autoref{fig:workflow} B3) which offers intelligent code completions (\autoref{fig:workflow} C1), real-time highlighting and preview (\autoref{fig:workflow} C2) upon invocation.

\CAMadd{Data contexts are integrated into the design of \textit{Xavier}.
Inspired by a previous study~\mbox{\cite{autoprofiler}}, we categorize data contexts into three types according to three different types of data objects~\mbox{\cite{table-scraps}}: tables, rows and columns.
Table-level data contexts contain basic information of a data table, including the table name, column name and shape.
Column-level data contexts include detailed information of a column.
Common column-level data contexts are the data type, null value count and sortedness.
In addition to common column-level contexts, categorical columns have unique values, value frequency, cardinality and value format, while numerical columns have value range, sample data points and value format.
Row-level data contexts include sample rows of a table, maintaining the spatial relationship between table cells.
Different data contexts can be combined and enhance functionalities in \textit{Xavier} like code completion.
For instance, column names from table-level contexts and sample rows from row-level contexts are both required for the table join operation to complete parameters indicating the key columns.
The same type of data contexts of different data objects can also be combined.
For example, table-level data contexts from multiple tables are essential for the table concatenation operation.}

\subsection{Data Context-Aware Code Completion}
\label{ssec:code-comp}

\begin{figure*}[ht]
  \centering
  \includegraphics[width=\linewidth]{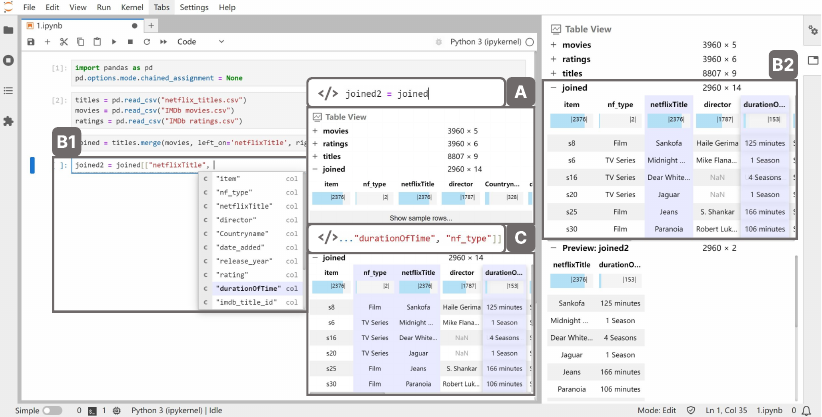}
  \caption{
    The usage scenario of automatic data context highlighting. A) \textit{Xavier} detected the existing DataFrame ``joined'' and showed the corresponding schema. B) When Sarah was selecting the suggested column names for the partial code (B1), \textit{Xavier} displayed sample rows of the DataFrame ``joined'' and highlighted relevant columns based on Sarah's code and the selected suggestion (B2). C) Finally, Sarah selected three columns which were highlighted by \textit{Xavier}.
  }
  \Description{An example to illustrate the effect of automatic highlighting of Xavier. Highlighted columns are marked as purple, which is the same color as that of the selected item in the selection list, since Xavier highlights the most relevant columns according to the user’s code and the selected item.}
  \label{fig:scenario-highlight}
\end{figure*}

\aptLtoX{\begin{shaded}
\noindent \DIFdel{Assume that Sarah, a data journalist, planned to analyze the relationship between the favourable rating and the duration of movies in a movie dataset.
She wanted to transform the dataset into a table with movie titles, ratings and duration before further analysis.
The dataset was complex, containing 3 tables and 20 columns, requiring custom wrangling code.
However, prior code completion tools offered poor suggestions as they did not have access to datasets fully leverage data contexts, making her frequently modify the suggested code, so she turned to Xavier for better code
completion coding assistance.}

\DIFadd{Assume that Sarah, a data journalist, planned to transform a movie dataset into a table with movie titles, ratings and duration to analyze the relationship between the favourable rating and the duration of movies.}
After loading \DIFdel{the 3}tables, Sarah found that the information about favourable ratings and the movie duration was not in the same DataFrame, so she decided to join them first.
As she wrote \lstinline{@joined@=@movies@.merge(@ratings@}\,, she forgot the exact spelling of the joined columns, though she remembered that they were about movie titles.
Without manually searching column names in the ``movies'' and ``ratings'' DataFrames, \textit{Xavier} suggested the completion \lstinline{@left_on@="netflixTitle",@right_on@="title"}.
Having a glimpse of corresponding highlight on the data view, Sarah found the code completion correct and immediately accepted the suggestion without manually typing verbose parameters.
\end{shaded}}{
\begingroup
\setlength{\fboxsep}{1em}
\noindent\colorbox{scenariocolor}{%
\begin{minipage}{\dimexpr\columnwidth-2\fboxsep\relax}
\setlength{\parindent}{1em} 
\setlength{\hangindent}{1em} 
\setlength{\hangafter}{1} 

\noindent \DIFdel{Assume that Sarah, a data journalist, planned to analyze the relationship between the favourable rating and the duration of movies in a movie dataset.
She wanted to transform the dataset into a table with movie titles, ratings and duration before further analysis.
The dataset was complex, containing 3 tables and 20 columns, requiring custom wrangling code.
However, prior code completion tools offered poor suggestions as they did not have access to datasets fully leverage data contexts, making her frequently modify the suggested code, so she turned to Xavier for better code
completion coding assistance.}

\DIFadd{Assume that Sarah, a data journalist, planned to transform a movie dataset into a table with movie titles, ratings and duration to analyze the relationship between the favourable rating and the duration of movies.}
After loading \DIFdel{the 3}tables, Sarah found that the information about favourable ratings and the
\end{minipage}%
}
\endgroup

\begingroup
\setlength{\fboxsep}{1em}
\noindent\colorbox{scenariocolor}{%
\begin{minipage}{\dimexpr\columnwidth-2\fboxsep\relax}
\setlength{\parindent}{1em} 
\setlength{\hangindent}{1em} 
\setlength{\hangafter}{1}
movie duration was not in the same DataFrame, so she decided to join them first.
As she wrote \lstinline{@joined@=@movies@.merge(@ratings@}\,, she forgot the exact spelling of the joined columns, though she remembered that they were about movie titles.
Without manually searching column names in the ``movies'' and ``ratings'' DataFrames, \textit{Xavier} suggested the completion \lstinline{@left_on@="netflixTitle",@right_on@="title"}.
Having a glimpse of corresponding highlight on the data view, Sarah found the code completion correct and immediately accepted the suggestion without typing verbose parameters.
\end{minipage}%
}
\endgroup}

\addvspace{0.7\baselineskip}

\textit{Xavier} incorporates both code contexts and data contexts to provide data context-aware code completion during typing (\textbf{R1}).
The functionality is implemented by a completion generator (\autoref{fig:workflow} B3) which organizes code contexts from \DIFadd{the} code context manager (\autoref{fig:workflow} B1) and data contexts from \DIFadd{the} data context manager (\autoref{fig:workflow} B2) to compute completion results.
The workflow consists of three steps, namely code context detection, data context organization, and code completion generation.

\subsubsection{Code context detection}
\label{sssec:code-context-detect}

\begin{figure*}[ht]
  \centering
  \includegraphics[width=\linewidth]{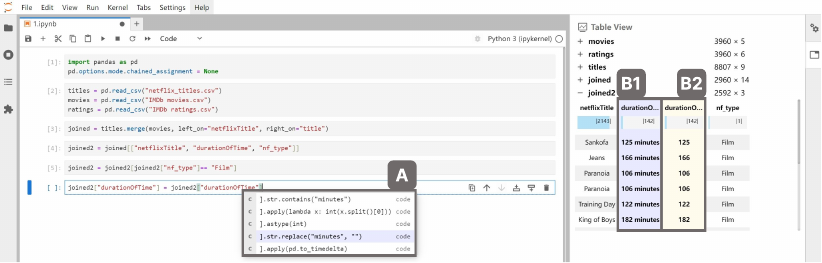}
  \caption{
    The usage scenario of real-time transformation preview. When Sarah switched to a completion item about column format transformation (A), \textit{Xavier} automatically computed the transformation result and added a preview column (B2) to the right of the original column (B1), with bold text in changed table cells.
  }
  \Description{An example to illustrate the effect of real-time transformation preview of Xavier. The preview column is marked as yellow, a different color to distinguish the preview column from the original one being transformed (marked as purple).}
  \label{fig:scenario-preview}
  \addvspace{-7pt}
\end{figure*}

Leveraging \DIFdel{the principle}\DIFadd{the principles} of lexical \DIFdel{analysis}and syntax analysis~\cite{dragon-book}, the code context manager (\autoref{fig:workflow} B1) detects the cursor position in the incomplete Python statement to determine missing parts of code.
A set of configurable and extensible rules is manually constructed based on the Python grammar~\cite{py-grammar}.
In the first step, the \DIFadd{code context} manager determines whether the \DIFadd{user's} input cursor is currently positioned within the signature of a Pandas function (e.g. \mycode{\lstinline{@pd@.merge(@df1@, @df2@, @left_on@="col1"}}) or out of the signature (e.g. \mycode{\lstinline{@df@[@df@["A"]}}).
If the cursor is in a signature, the manager records the signature name and \DIFadd{identifies} missing parameters in the partial code.
For example, \DIFdel{for}\DIFadd{in} the partial code \mycode{\lstinline{@pd@.merge(@df1@, @df2@, @left_on@="col1"}}, the signature name is ``merge'' and the filled parameters are the left table ``df1'', the right table ``df2'' and the joined column ``col1'' in the left table.
\DIFdel{, which means a}\DIFadd{This indicates that a corresponding} joined column name in the right table is required.
If the cursor is out of the signature, the manager \DIFdel{determines the AST patterns}\DIFadd{analyzes patterns of Abstract Syntax Trees (ASTs)} to identify possible transformation operators and missing AST nodes.
For instance, \DIFdel{for}\DIFadd{in} the partial code \mycode{\lstinline{@df@[@df@["A"]}}, the manager identifies it as a filtering operation according to the AST pattern \mycode{\lstinline{@df@[}}.
Based on the basic format of a filtering condition, at least an operator and corresponding parameters like \mycode{\lstinline{== "Value"}} are required to complete the \DIFdel{filtering condition}\DIFadd{transformation}.
\DIFdel{According to Python grammar~\mbox{\cite{py-grammar}} and Pandas documentation~\mbox{\cite{doc-pandas}}, the code context manager only considers the innermost incomplete transformation or the last incomplete transformation for generalizability.}%
\DIFadd{To ensure generalizability, the code context manager matches AST patterns from child nodes up to parent nodes.}
Hence, for nested transformations like \mycode{\lstinline{@pd@.merge(@df1@, @df2@[@df2@["A"]}}, only the filtering transformation \mycode{\lstinline{@df2@[@df2@["A"]}} will be considered. 
Likewise, for chained transformations like \mycode{\lstinline{@df1@["A"].fillna("Unknown").~str~.replace(}}, only the string \DIFdel{replace}\DIFadd{replacement} transformation function \mycode{\lstinline{replace(}} will be taken into account.
These code contexts will be utilized to determine the\DIFdel{optimal} timing for triggering \DIFadd{each category of} code completion suggestions \DIFadd{(see Table 1 and Table 2 in the appendix)} and to inform the content of those suggestions.

\subsubsection{Data context organization}
\label{sssec:data-context-organize}

As discussed in Section ~\ref{ssec:overview}, three types of data contexts in the data context manager (\autoref{fig:workflow} B2) have been pre-calculated since the last run and will be sent to the completion generator (\autoref{fig:workflow} B3) as the code completion is triggered.
To save the computation cost and offer more targeted completions, the data \DIFdel{context}\DIFadd{contexts} will be further filtered in two steps, namely type matching by operators and context selection by operands.

In type matching, data context types are determined based on transformation types classified in Pandas documentation~\cite{doc-pandas}.
\DIFadd{This step}\DIFdel{which} excludes data context types that have weak relevance \DIFdel{with}\DIFadd{to} the transformation.
Since ``DataFrame'' and ``Series'' cover most of commonly used operators in Pandas, we classify all operators into three types: ``DataFrame'', ``Series'', and ``Others''.
``DataFrame'' operators primarily rely on the table-level data contexts, with row-level and column-level data contexts as additional references.
For instance, \DIFdel{the semantics of}column names and sample values \DIFdel{are both the possible}\DIFadd{can provide useful} clues to complete \DIFadd{a} joined table \mycode{\lstinline{@country_code@}} and joined columns \mycode{\lstinline{@left_on@="Country", @right_on@="Countryname"}} \DIFdel{for}\DIFadd{in the} partial code \mycode{\lstinline{@df3@ = @covid_data@.merge(}}.
For ``Series'' operators, only column-level data contexts \DIFdel{will be}\DIFadd{are} selected, since \DIFdel{such}\DIFadd{``Series''} operators \DIFdel{primarily} focus on single or multiple columns in a table.
For example, the format of values in \DIFadd{the} Series \mycode{\lstinline{@df@["Country"]}} is an essential clue to \DIFdel{complete}\DIFadd{completing} substrings to be replaced in \mycode{\lstinline{@df@["Country"].~str~.replace(}}.
For ``Others'' operators, \DIFdel{only table-level data contexts are considered as fundamental data contexts.}\DIFadd{the data contexts considered are default to table-level data contexts.}

In context selection, data contexts of specific tables or columns are determined by missing parts of code detected in Section~\ref{sssec:code-context-detect}.
\DIFdel{which}\DIFadd{This step} further excludes weakly relevant data contexts \DIFdel{of}\DIFadd{from} other tables or columns.
For instance, in the partial code of a table filtering operation \mycode{\lstinline{@df@[}}, row-level data contexts like sample rows of ``df'' \DIFdel{will be}\DIFadd{are} taken into account as a supplementary clue in addition to table-level data contexts, since the filtering condition is totally unknown.
However, when completing the partial code \mycode{\lstinline{@df@[@df@["A"]}}, column-level data contexts, such as sample values of column ``A'', will take precedence over row-level data contexts, as the filtering condition is likely related to column ``A''.
As a result, within the same operator, the required data contexts will be dynamically adjusted according to the missing parts of the code, retaining only the most relevant data contexts to provide targeted code completion \DIFadd{(see Table 1 and Table 2 in the appendix for details).}
Since the \DIFadd{amount of} filtered data contexts can be potentially large (e.g. unique values of primary key columns), the \DIFdel{completion generator}\DIFadd{data context manager} randomly samples values to control the size of data contexts.
\DIFadd{To reduce the sampling limitation, the data context manager also supports filtering data values by prefix, saving the effort of manually adding information about data values into comments before generating code.}
For instance, \DIFadd{although} at most 50 unique values are sampled for the ``unique values'' data context of a categorical column, \DIFadd{users can still specify data values by typing the prefix of values.}

\subsubsection{Code completion generation}

With the code context detected and data context organized, \textit{Xavier} combines both \DIFdel{code context and data context} to offer recommendations \DIFdel{of}\DIFadd{for} subsequent code through the completion generator (\autoref{fig:workflow} B3), supporting \DIFadd{both} single-token and multi-token completion.
The single-token completion works like traditional code completion \DIFdel{to complete}\DIFadd{by completing} the rest of a token \DIFdel{by prefixes}\DIFadd{based on its prefix}, but it provides more comprehensive column name completions since writing column names is common in data wrangling scripts.
For instance, for the partial code \mycode{\lstinline{@df@.sort_values(@by@="C}}, all \DIFdel{of the} column names starting with ``C'' in table ``df'' will be listed \DIFadd{as suggestions}.
The multi-token completion leverages the capabilities of Llama3-70B~\cite{llama3}, an open-source LLM with outstanding performance in various general tasks.
Inspired by the prompt structure \DIFadd{proposed} in~\cite{nl2rigel}, we \DIFdel{have designed}\DIFadd{design} a prompt template consisting of four components: \textit{code context}, \textit{data context}, \textit{task instruction} and \textit{format control}.
The \textit{code context} refers to all data wrangling code preceding the cursor.
The \textit{data context} is the textual representation of the organized data \DIFdel{context}\DIFadd{contexts} discussed in Section~\ref{sssec:data-context-organize}.
\DIFdel{, with}Column-level data contexts \DIFdel{organized}\DIFadd{are grouped} by columns\DIFadd{, while}\DIFdel{and} table-level \DIFdel{or}\DIFadd{and} row-level data contexts \DIFadd{are} grouped by tables.
The \textit{task instruction} includes the thinking steps to help derive the appropriate completions.
\textit{Format control} is employed to restrict the output format, increasing the likelihood of obtaining syntactically correct answers.
Note that the prompt template is merely one of the feasible solutions to \DIFdel{combine}\DIFadd{combining} code and data contexts to offer data context-aware code completion\DIFadd{. We}\DIFdel{and we} encourage future researchers to explore alternative solutions. 

\subsection{Automatic Data Context Highlighting}
\label{ssec:data-ctx-highlight}

\begin{figure*}[t]
  \centering
  \includegraphics[width=\linewidth]{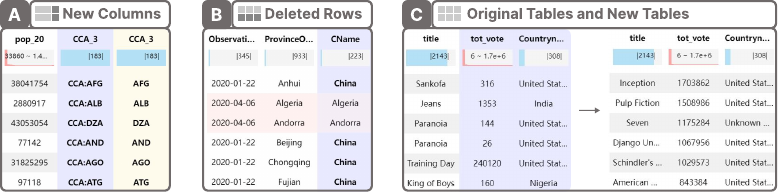}
  \caption[]{
    Three preview forms of \textit{Xavier}. A) For the column format transformation, a new column is created to the right of the original column. B) For the table filtering transformation, rows to be deleted are highlighted. C) For transformations that generate a new table or change the whole table \CAMdel{(e.g. summing up cases for each country when there are multiple provinces or states involved)}\CAMadd{(e.g. Sort movies by total votes. For movies having equal total votes, sort them by country names)}, both the original table and the result table are displayed.
  }
  \Description{Three preview forms of Xavier in a spreadsheet style.}
  \label{fig:highlightform}
  \addvspace{-7pt}
\end{figure*}

\aptLtoX{\begin{shaded}
\noindent 
\DIFdel{Having written code of the table join}\DIFadd{After writing the code to join tables}, Sarah ran \DIFdel{the code}\DIFadd{it} to compute the \lstinline{@joined@} DataFrame.
\textit{Xavier} detected the change of the notebook memory and added profiles of the \lstinline{@joined@} DataFrame to the data view with schemas automatically displayed, saving the need to manually write code to print the result DataFrame.
Glancing at the schema, Sarah found that the \lstinline{@joined@} DataFrame included 14 columns, with too many unrelated ones (\autoref{fig:scenario-highlight} A).
She decided to select the related ones first for convenience of subsequent wrangling.

As she wrote \lstinline{@joined2@=@joined@}\,, \textit{Xavier}\DIFdel{automatically} folded the \DIFdel{schema}\DIFadd{schemas} of other tables, making her focus on the current table (\autoref{fig:scenario-highlight} A).
\DIFadd{While} selecting the suggested column names for the partial code like \lstinline{@joined2@ \ = @joined@[["netflixTitle"} (\autoref{fig:scenario-highlight} B1), \textit{Xavier} \DIFdel{automatically}showed sample rows of the DataFrame ``joined'' and highlighted columns mentioned in the partial code (i.e. the column ``netflixTitle'') \DIFdel{and}\DIFadd{as well as} the \DIFadd{currently} focused completion item ``durationOfTime'', as is shown in \autoref{fig:scenario-highlight} B2.
Since the data view had limited width and could not display all columns of the DataFrame at the same time, \textit{Xavier} anchored invisible highlighted columns (i.e. ``durationOfTime'' in \autoref{fig:scenario-highlight} B2) for Sarah to \DIFdel{refer to}\DIFadd{reference}.
\DIFdel{Therefore, Sarah could}\DIFadd{This feature allowed her to} review the selected columns while switching among different completion items, \DIFdel{to ensure}\DIFadd{ensuring that} these columns were what she wanted.
Finally, Sarah typed the code \lstinline{@joined2@ \ = @joined@[["netflixTitle","durationOfTime","nf_type"]]} (\autoref{fig:scenario-highlight} C) and executed it with confidence.
\end{shaded}}{
\begingroup
\setlength{\fboxsep}{1em}
\noindent\colorbox{scenariocolor}{%
\begin{minipage}{\dimexpr\columnwidth-2\fboxsep\relax}
\setlength{\parindent}{1em} 
\setlength{\hangindent}{1em} 
\setlength{\hangafter}{1} 

\noindent 
\DIFdel{Having written code of the table join}\DIFadd{After writing the code to join tables}, Sarah ran \DIFdel{the code}\DIFadd{it} to compute the \lstinline{@joined@} DataFrame.
\textit{Xavier} detected the change of the notebook memory and added profiles of the \lstinline{@joined@} DataFrame to the data view with schemas automatically displayed, saving the need to manually write code to print the result DataFrame.
Glancing at the schema, Sarah found that the \lstinline{@joined@} DataFrame included 14 columns, with too many unrelated ones (\autoref{fig:scenario-highlight} A).
She decided to select the related ones first for convenience of subsequent wrangling.

As she wrote \lstinline{@joined2@=@joined@}\,, \textit{Xavier}\DIFdel{automatically} folded the \DIFdel{schema}\DIFadd{schemas} of other tables, making her focus on the current table (\autoref{fig:scenario-highlight} A).
\DIFadd{While} selecting the suggested column names for the partial code like \lstinline{@joined2@ \ = @joined@[["netflixTitle"} (\autoref{fig:scenario-highlight} B1), \textit{Xavier} \DIFdel{automatically}showed sample rows of the DataFrame ``joined'' and highlighted columns mentioned in the partial code (i.e. the column ``netflixTitle'') \DIFdel{and}\DIFadd{as well as} the \DIFadd{currently} focused completion item ``durationOfTime'', as is shown in \autoref{fig:scenario-highlight} B2.
Since the data view had limited width and could not display all columns of the DataFrame at the same time, \textit{Xavier} anchored invisible highlighted columns (i.e. ``durationOfTime'' in \autoref{fig:scenario-highlight} B2) for Sarah to \DIFdel{refer to}\DIFadd{reference}.
\DIFdel{Therefore, Sarah could}\DIFadd{This feature allowed her to} review the selected columns while switching among different completion items, \DIFdel{to ensure}\DIFadd{ensuring that} these columns were what she wanted.
Finally, Sarah typed the code \lstinline{@joined2@ \ = @joined@[["netflixTitle","durationOfTime","nf_type"]]} (\autoref{fig:scenario-highlight} C) and executed it with confidence.
\end{minipage}%
}
\endgroup}

\addvspace{0.4\baselineskip}

In order to help users rapidly locate relevant parts of datasets (\textbf{R2}), \textit{Xavier} automatically unfolds profiles of the corresponding DataFrames and highlights columns in the side panel when users are typing code or switching between different completion items.
\textit{Xavier} detects table and column names mentioned in the partial code \DIFdel{and the completion items being focused}\DIFadd{as well as those in the currently focused completion item}.
When the partial code and completion item only contain table names, \textit{Xavier} simply displays the schema of each table, with sample rows concealed.
For example, when users are filling in the table names in a union transformation, such as \mycode{\lstinline{@pd@.concat([@df1@}}, and the currently focused completion item is \mycode{\lstinline{@df2@, @df3@]}}, the schemas of tables ``df1'', ``df2'', and ``df3'' will be automatically presented in the side panel, while other previously unfolded table profiles are collapsed.
When column names are also present, sample rows of tables will be shown and the mentioned columns will be highlighted.
For instance, as users are selecting columns of a table through \mycode{\lstinline{@df@[["col1", "col2",}} and \DIFadd{the} currently focused completion item is \mycode{\lstinline{"col3"}}, all \DIFdel{the} three columns \DIFadd{(``col1'', ``col2'' and ``col3'')} will be marked using a different background color in the sample rows.
Since the width of the side panel is limited and cannot simultaneously display all columns in a table, \textit{Xavier} adopts a floating effect to anchor  highlighted columns to the right side of the data view.
Therefore, users do not have to frequently scroll horizontally across the sample row view to find all highlighted columns.
Automatic data context highlight helps users rapidly recall tables or columns of interest, keeping users aware of data contexts in the current transformation although the code is not necessarily complete.

\subsection{Real-time Transformation Preview}
\label{ssec:data-cxt-preview}

\aptLtoX{\begin{shaded}
\noindent 
Having selected relevant columns and obtained a new DataFrame \lstinline{@joined2@}, Sarah wanted to standardize the format of the ``durationOfTime'' column in order to better analyze the relationship between the favourable rating and the duration of movies.
When she wrote \lstinline{@joined2@["durationOfTime"] = @joined2@["durationOfTime"]}, \textit{Xavier} suggested multiple wrangling operations, including \lstinline{.~str~.replace(" minutes", "")} (\autoref{fig:scenario-preview} A).
Not sure about the effect of the operation, Sarah switched to \DIFdel{that}\DIFadd{the} completion item and checked the preview in the data view.
She found that a new column marked in yellow (\autoref{fig:scenario-preview} B2) was positioned beside the ``durationOfTime'' column (\autoref{fig:scenario-preview} B1), clearly showing the \DIFdel{transformation preview}\DIFadd{changes} with bold text in \DIFdel{changed}\DIFadd{modified} table cells.
Sarah immediately verified the suggestion and accepted it to standardize the column \DIFadd{format} with ease.
\end{shaded}}{
\begingroup
\setlength{\fboxsep}{1em}
\noindent\colorbox{scenariocolor}{%
\begin{minipage}{\dimexpr\columnwidth-2\fboxsep\relax}
\setlength{\parindent}{1em} 
\setlength{\hangindent}{1em} 
\setlength{\hangafter}{1} 

\noindent 
Having selected relevant columns and obtained a new DataFrame \lstinline{@joined2@}, Sarah wanted to standardize the format of the ``durationOfTime'' column in order to better analyze the relationship between the favourable rating and the duration of movies.
When she wrote \lstinline{@joined2@["durationOfTime"] = @joined2@["durationOfTime"]}, \textit{Xavier} suggested multiple wrangling operations, including \lstinline{.~str~.replace(" minutes", "")} (\autoref{fig:scenario-preview} A).
Not sure about the effect of the operation, Sarah switched to \DIFdel{that}\DIFadd{the} completion item and checked the preview in the data view.
She found that a new column marked in yellow (\autoref{fig:scenario-preview} B2)
\end{minipage}%
}
\endgroup

\begingroup
\setlength{\fboxsep}{1em}
\noindent\colorbox{scenariocolor}{%
\begin{minipage}{\dimexpr\columnwidth-2\fboxsep\relax}
\setlength{\parindent}{1em} 
\setlength{\hangindent}{1em} 
\setlength{\hangafter}{1}
was positioned beside the ``durationOfTime'' column (\autoref{fig:scenario-preview} B1), showing the \DIFdel{transformation preview}\DIFadd{changes} with bold text in \DIFdel{changed}\DIFadd{modified} table cells.
Sarah immediately verified the suggestion and accepted it to standardize the column \DIFadd{format} with ease.
\end{minipage}%
}
\endgroup}

\addvspace{0.4\baselineskip}

In order to facilitate understanding and straightforward verification of code completions (\textbf{R3}), \textit{Xavier} provides preview in \DIFadd{the} data view for users to immediately examine the transformation results when required parameters for an operation are all filled, without explicit code execution by users.
Inspired by Wrangler~\cite{wrangler}, \textit{Xavier} supports three types of \DIFdel{preview}\DIFadd{previews}, as is shown in \autoref{fig:highlightform}.
For column format transformation like substring replacement and null value filling, a new column marked \DIFdel{as}\DIFadd{in} yellow is created to the right of the original column, with bold text \DIFdel{in}\DIFadd{indicating the} changed cells (\autoref{fig:highlightform} A).
For table filtering transformation, deleted rows \DIFdel{will be marked as}\DIFadd{are marked in} red, with bold text of filtering values mentioned in user's code (\autoref{fig:highlightform} B).
For transformations that generate a new result table or change the whole original table like table sorting, aggregation and merging, \textit{Xavier} displays both the original table and the new table (\autoref{fig:highlightform} C).
\DIFadd{The preview visualization is similar to the design of TweakIt~\cite{tweakit}, although \textit{Xavier}'s preview is intended for comprehending AI-generated code completions instead of pre-existing code snippets.
}

\section{User Study}
\label{sec:user-study}
To evaluate the effectiveness and usability of \textit{Xavier}, we conducted a comparative user study\footnote{The user study has been approved by State Key Lab of CAD\&CG, Zhejiang University.} in which participants were asked to complete data wrangling tasks using \textit{Xavier} and a baseline tool. 
The setup of our user study is delineated in Sections~\ref{sec:us-participants}-\ref{sec:us-procedure} and the results are reported in Section~\ref{sec:us-results}.

\subsection{Participants}
\label{sec:us-participants}
In our user study, 16 data analysts (denoted as U1-U16, 11 male and 5 female, $Age_{mean}=23.94$, $Age_{std}=1.69$, $Experience_{mean}=3.06\ years$, $Experience_{std}=1.77\ years$) were recruited from a university through social media and word-of-mouth.
\DIFadd{This was a completely fresh sample with no overlap between participants in this study and those involved in the preliminary study.}
They had diverse backgrounds including Software Engineering, Energy, Management Science and Engineering, Data Visualization, Physical Education and Training, and Computer Science.
They all had programming experience in data wrangling ranging from one year to six years and regularly used Python Pandas in computational notebooks.
Participants consented to having their voices and programming processes recorded.

\subsection{Apparatus and Materials}
\label{sec:us-task_and_dataset}

\hspace*{\parindent}\textbf{Apparatus.}
The baseline tool we constructed was different from \textit{Xavier} in two aspects: code completion and visualization view.
For code completion, the baseline tool used the same backend LLM as that of \textit{Xavier}, while data contexts were removed from the prompt to compare the effectiveness of data contexts in code suggestions.
\DIFdel{Theoretically, the underlying LLM of both the baseline tool and \textit{Xavier} should be the model of GitHub Copilot, but Copilot did not provide a public API to directly customize the prompt behind code completion.
To compare code completion with and without data contexts under the same model condition,}%
We chose Llama3-70B~\cite{llama3}, an exemplary publicly-available LLM as the underlying model for both the baseline tool and \textit{Xavier}, \DIFadd{due to its relatively low latency in local deployment}.
For the fairness of tool comparison, the side panel view of AutoProfiler~\cite{autoprofiler}, a continuous data profiling tool in recent literature, was replicated for the visualization view.
Similar to Section~\ref{ssec:preliminary-apparatus-material}, we also prepared a slide with task descriptions and data dictionaries.
In order to control the standardization of participants' data transformation scripts and simultaneously reduce the cost of participants referring to APIs, we prepared a cheatsheet for them.
Inspired by~\cite{cheatsheet-pandas}, the cheatsheet listed supported transformations by \textit{Xavier} and corresponding API usage, categorizing APIs by functionality.

\textbf{Datasets}.
We crafted two datasets, denoted as \textit{Covid-19} and \textit{Movies} dataset from two notebooks\footnote{\url{https://www.kaggle.com/code/erikbruin/storytelling-covid-19} and \url{https://www.kaggle.com/code/niharika41298/netflix-visualizations-recommendation-eda}} on Kaggle.
We chose these notebooks since participants are generally familiar with the background of datasets~\cite{task-based-effect-vis} and typically do not spend much time understanding the data.\DIFdel{To control and balance the difficulty of the tasks, we adjusted the number and size of tables, keeping 3 tables and 20 columns for each dataset.}
The crafted datasets still covered two major types of tabular data (i.e. categorical and numerical).
For the similar reason discussed in Section~\ref{ssec:preliminary-apparatus-material}, data tables were slightly modified.

\textbf{Tasks}.
We designed a data wrangling task \DIFdel{containing approximately 10 lines of code} for each dataset in a similar way to that described in Section~\ref{ssec:preliminary-apparatus-material}.
\DIFadd{\CAMdel{To control and balance the difficulty of the tasks,}\CAMadd{To avoid users being distracted by too much irrelevant data and spending much time exploring the dataset,} we controlled the number of tables, columns, types of data transformation operators, and the number of data transformation steps involved in both tasks.
For each dataset, we only selected 3 tables and 20 columns relevant to the corresponding task.
\CAMadd{In making these selections, we sought to preserve the logical and semantic relationships within the original dataset as much as possible, such as retaining primary and foreign key relationships.
This ensures that participants can complete the tasks using only the provided data without requiring additional information.}
Both tasks share the same set of data transformation operators.
Each task contains approximately 10 lines of code.}
For each task, participants were asked to construct a table with 4 attributes.
In the remaining part of Section~\ref{sec:user-study}, \textit{Movie Task} refers to the data wrangling task on \textit{Movies} dataset, and \textit{Covid Task} refers to the one on \textit{Covid-19} dataset.
Task details are left to the supplementary materials.

\subsection{Procedure}
\label{sec:us-procedure}
We opted for a counterbalanced mixed design to compare \textit{Xavier} and the baseline tool.
We denote the two systems as $X$(avier) and $B$(aseline) and the two datasets as $M$(ovies) and $C$(ovid-19).
The participants were divided into groups of four.
In each group, the participants covered the following four experimental conditions: $[MB, CX], [MX, CB], [CB, MX], [CX, MB]$.
Such approach allowed each participant to experience both tools and reduced potential impact on experiment data brought by the sequence of tool using and tasks like learning effect.

Like Section~\ref{ssec:pre-study-procedure}, we informed participants about relevant information, conducted a code authoring experiment (50-60 minutes) and a semi-structured interview (10-15 minutes), and collected data from recordings.
The entire study took around 75 minutes and each participant received 70 Chinese Yuan as compensation.

\textbf{Code authoring experiment}.
Similar to Section~\ref{ssec:pre-study-procedure}, we initially introduced the tool before each task and allowed participants to try it on a warm-up task.
To verify whether introducing data context can improve code completion and enhance user experience, we did not inform participants in advance about how these two tools worked.
When participants were ready, they started to author data wrangling scripts assisted by \textit{Xavier} or the baseline tool.
Participants were permitted to consult the cheatsheet, the data dictionary and the task description at any time during authoring.
After the experiment, they were asked to finish a questionnaire which assessed their perceived workload.

\textbf{Semi-structured interview}.
Three parts consisted of the semi-structured interview.
Initially, we asked participants to compare the effect of code completion between the two tools corresponding to \textbf{R1} in Section~\ref{ssec:user-requirements}.
Then, participants compared and discussed the coding assistance brought by the side panel, which corresponds to \textbf{R2} and \textbf{R3}.
In the last part, we asked questions specifically targeting the issues observed during the experiment.
The questionnaire included six NASA-TLX~\cite{nasa-tlx} questions to measure the perceived workload of participants when using Baseline and \textit{Xavier}.
All questions were measured using the 7-point Likert Scale.

\subsection{Results}
\label{sec:us-results}

\begin{figure}[t]
  \centering
  \includegraphics[width=\columnwidth]{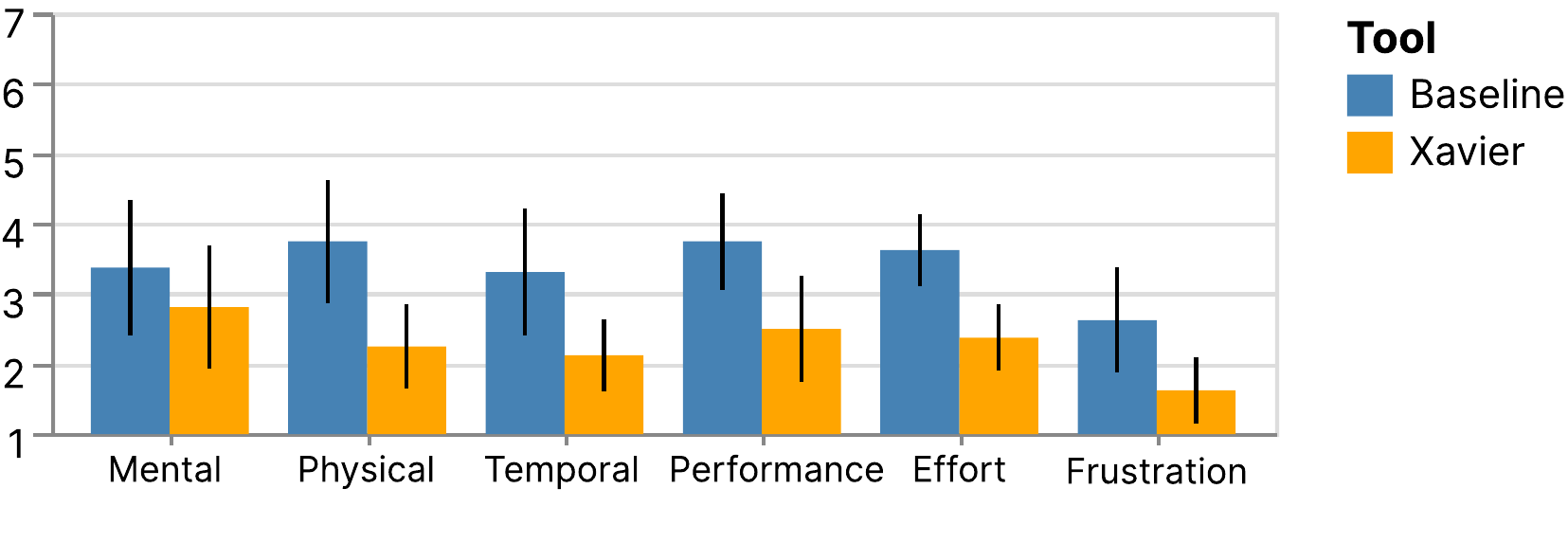}
  \caption{
    The perceived workload of participants using the baseline tool and \textit{Xavier}.
  }
  \Description{A bar chart with two groups: Xavier and baseline. The x axis represents different questions in NASA-TLX questionnaire: mental, physical, temporal, performance, effort, frustration. The y axis represents the score ranging from 1 to 7.}
  \label{fig:questionnaire}
\end{figure}

\begin{figure}[t]
  \centering
  \includegraphics[width=0.7\columnwidth]{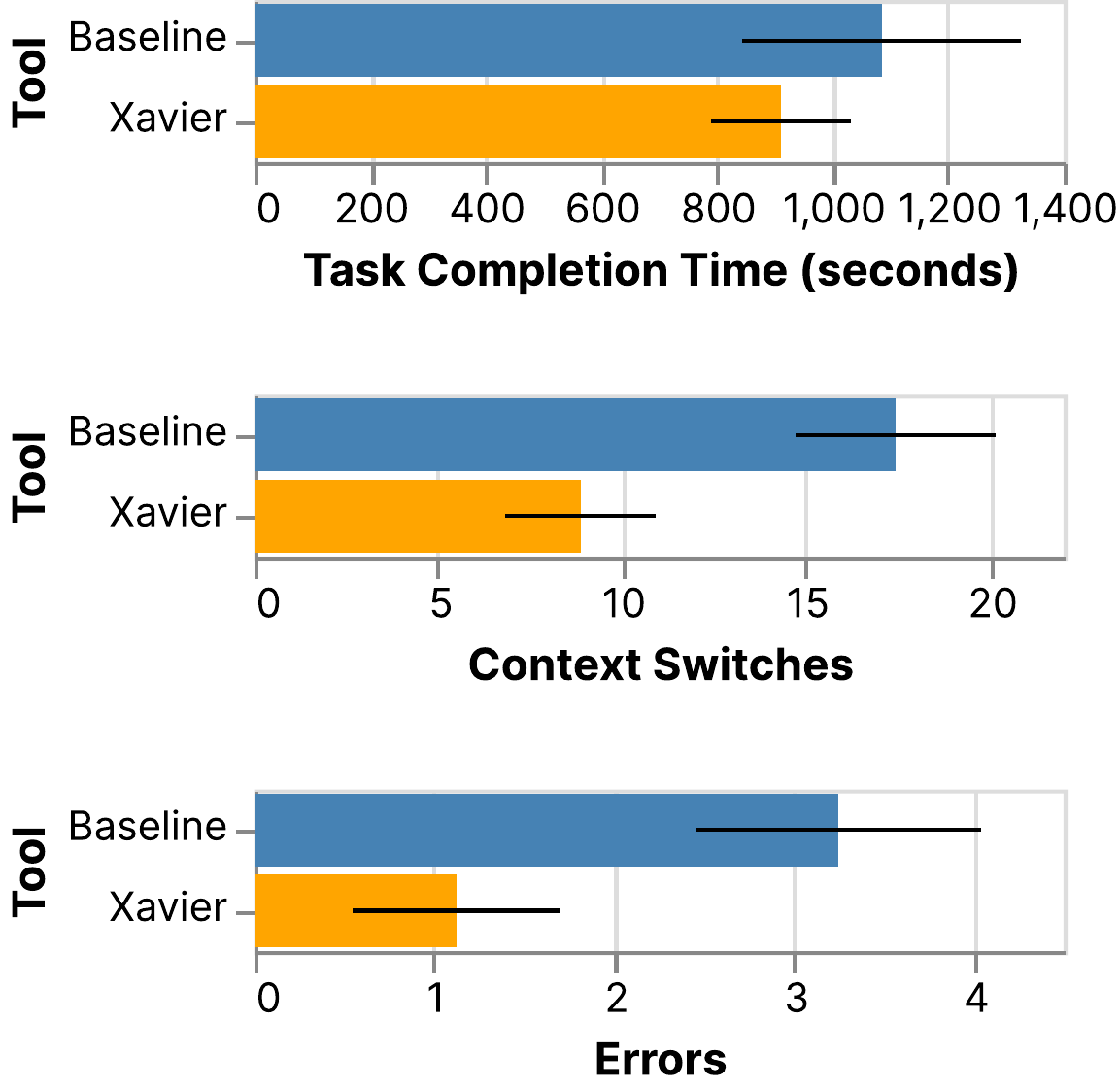}
  \caption{
    Quantitative results of our user study. We recorded task completion time in seconds, context switches and errors of participants by using the baseline tool and \textit{Xavier}.
  }
  \Description{There are three bar charts. The first chart is about the task completion time by using the two tools (Baseline: 1085.875 seconds, Xavier: 911.563 seconds). The second chart is about context switches (Baseline: 17.438, Xavier: 8.875). The third chart is about errors (Baseline: 3.250, Xavier: 1.125).}
  \label{fig:userstudyquan}
\end{figure}

\subsubsection{Quantitative Results}
\label{sssec:quantitative-res}

We collected and summarized the questionnaire results shown in \autoref{fig:questionnaire}.
On average, participants perceived relatively lower workload when using \textit{Xavier} compared to that of the baseline tool regarding six aspects (mental demand, physical demand, temporal demand, performance, effort and frustration), which suggests that the data context-aware design of \textit{Xavier} has the potential to reduce workload in code authoring.

To figure out whether \textit{Xavier} accelerated the coding process, we measured ``task completion time'' of each participant for each \DIFdel{case}\DIFadd{task}.
To assess whether \textit{Xavier} facilitated locating and understanding relevant data (\textbf{R2} and \textbf{R3}), we defined ``context switches'' to estimate the additional overhead of authoring data wrangling scripts.
A context switch starts when users pause writing data wrangling scripts, during which users may manipulate the side panel by mouse, manually add code to profile the data, or leave the current editor window to review the original data tables.
The context switch ends when users return to the editor and continue writing the wrangling scripts.
To evaluate the suggestion correctness of \textit{Xavier}, we defined ``errors'' to measure the errors users encountered in the code authoring experiment.
The errors include code errors related to language grammar and data errors like accessing non-existent columns, unexpected transformation results and so on.

Since the raw data of these measurements did not obey either normal distribution or equal variance, we used
the Scheirer-Ray-Hare test, a nonparametric substitute for ANOVA, for statistical tests.
We tested the main effects of tools and \DIFdel{cases}\DIFadd{tasks}, as well as their interaction effects. 
We found that for all three measurements, the interaction effects and the main effects of \DIFdel{case}\DIFadd{tasks} are not significant.
For task completion time, the main effect of tools is not significant.
However, for context switches and errors, the main effect of tools is significant (both $p < 0.001$). 
Considering that the interaction effects and the main effects of tool are not significant, we reported the means of task completion time, context switches, and errors of different tools, as shown in \autoref{fig:userstudyquan}.
There are fewer context switches and errors when users using \textit{Xavier}, which indicates that participants likely concentrated on data wrangling code authoring more easily and wrote data wrangling code with more accuracy with \textit{Xavier}.
However, we did not find a significant main effect of tools on task completion time. 
One of the possible reasons is that the task design is relatively simple (around 10 lines of code for each task) given the limited experimental duration.
Hence, long-term evaluation of the task completion time by using \textit{Xavier} is required in the future work.

\subsubsection{Qualitative Results}
\label{sssec:qualitative-res}
\DIFadd{To analyze the user feedback from the interview, we conducted a qualitative inductive content analysis~\cite{content-analysis}.
We focused on three aspects: 1) how the participants compared and evaluated the two tools; 2) the comments of the participants on \textit{Xavier}; 3) the issues they encountered when using Xavier and their suggestions for improvement.
In the open coding phase, two co-authors (data analysis experience: $\geq 3\ years$) read the recording transcripts and labeled the interview feedback individually.
\CAMdel{Then they collaboratively discussed the labeled content and grouped the similar comments.}\CAMadd{Then they collaboratively resolved disagreements through periodic meetings where the coders discussed the labeled content and reconciled differences in interpretation.
Furthermore, they grouped the similar comments into higher-level themes.}
\CAMdel{Furthermore,}\CAMadd{Finally,} an additional co-author (data analysis experience: $> 5\ years$) was involved to derive the consensus on the key findings, which are as follows: 
}

\DIFadd{\textbf{User experience.} 
Most participants (14/16) thought that \textit{Xavier}’s code completions were ``\textit{better}''.
Some participants even described the code completions of \textit{Xavier} as ``\textit{amazing}'' (U4), ``\textit{useful}'' (U1, U6), or ``\textit{smart}'' (U5).
Similarly, most participants (14/16) mentioned the benefits of the automatic preview of \textit{Xavier}, thinking that it is ``\textit{helpful}'' (U1, U3, U12) and it can help them ``\textit{avoid errors}'' (U5, U8, U14).
The third feature of \textit{Xavier} frequently mentioned by participants is automatic highlighting (13/16).
Participants described the automatic highlighting as ``\textit{intuitive}'' (U11, U15) and ``\textit{convenient}'' (U3, U5, U14).
As the live data view kept changing while coding, we additionally asked participants if they were distracted by frequent changes.
However, none of the participants found it distracting. 
We identified two primary reasons from their responses.
First, their attention primarily remained on the code editor during typing and they only saw the live data view when necessary (U2-U4, U6, U10, U11, U13, U16).
Second, real-time updates of the live data view can bring them confidence (U1, U3, U8, U15).
For example, ``\textit{If it (the live data view) doesn’t change, I will worry if I have written something wrong}'', U8 noted.}

\DIFadd{\textbf{Code completion preferences}.
Participants seemed to prefer shorter code completions rather than longer ones.
According to the participants' feedback, they preferred completions like column names (U1-U5, U8, U11, U13-U16), data values (U3, U16) and parameters (U8, U13, U14), rather than completions like the whole transformation statement.
We identified two primary reasons for the preference of shorter code completions. 
First, longer completions typically required more time to generate (U2, U6, U12).
For instance, U6 skipped most of the completions (8/10) of the whole statement after the assignment operator.
As she pointed out, ``\textit{It (the code completion of Xavier) came out a little slower than I thought, making me hesitate whether to type it myself or wait for it to appear}''.
Second, participants believed longer completions would be less accurate (U12-U15).
As U14 commented, ``\textit{If it (Xavier) could really guess what I'm thinking and provide longer completions, that would be the best.
But I probably don't have such expectations because I think it's very difficult to achieve.}''}

\DIFadd{\textbf{Transparency and trust}.
As previously discussed, the preview and highlight features of \textit{Xavier} may help users more easily validate the completions and enhance their confidence in the tool.
However, these features cannot fully address the issues of transparency and trust.
One issue is that the model mechanisms behind completion are not transparent.
Specifically, U4, U6, U15 and U16 were curious about how the code completion is recommended.
For instance, U4 was surprised by the accuracy of the completion \mycode{\lstinline{@df@["ConfirmedCases"] \ \ / @df@["pop_20"]}} for the partial code \mycode{\lstinline{@df@["casesPer"] \ =}}, which was exactly what he wanted.
``\textit{How did it (Xavier) know the meaning of `pop\_20' is the population in 2020?
Is there some information to tell it (Xavier)?}'', U4 asked.
We also noticed that initial few code completions may influence participants' trust in its capabilities (U8, U11, U13, U15).
For instance, during the code authoring experiment, U15 initially waited for the completion of the whole filtering transformation when she typed \mycode{\lstinline{@selected@ \ =}}, but did not receive a correct response.
Eventually, she skipped most of the completions of this kind (6/7) and only waited for shorter ones as she specified the transformation operator.
``\textit{Sometimes it (Xavier) completed a long line for me... I would not trust such completions}'', she said in the interview.
}

\section{Discussion}

This paper presents \textit{Xavier}, a tool designed to enhance data wrangling script authoring in computational notebooks.
\DIFadd{The user study revealed that users encountered significantly fewer context switches and errors during scripting by using \textit{Xavier} (Section~\ref{sssec:quantitative-res}). 
User feedback indicated that \textit{Xavier} could help validate code and bring confidence to users (Section~\ref{sssec:qualitative-res}).
In this section, we discuss the lessons learned from the development and evaluation of \textit{Xavier} in Section~\ref{ssec:lessons-learned}.
Besides, we identify limitations of our research that can be further improved in Section~\ref{ssec:limitation-future}.
}

\DIFdel{We identify two key implications:}

\DIFdel{\textbf{Enhancing coding assistance with data contexts in data wrangling.}
In this paper, we introduce data contexts for both \textit{Xavier} and users.
With data contexts integration, \textit{Xavier} has the potential to provide more intelligent code completion.
Automatic highlighting and preview keeps users aware of the data contexts throughout the coding process, facilitating location of relevant parts of data and easy verification.
By incorporating data contexts for both coding assistant and users, \textit{Xavier} streamlines the process of data wrangling programming.}

\DIFdel{\textbf{Improving user experience by instant feedback.}
When completing data wrangling tasks, users often need to verify the correctness of the code, commonly through data profiling. Especially in scenarios where AI is used to generate code, users tend to spend more time on code verification and modification rather than writing code. Immediate feedback can enable users to complete code verification more easily, while also enhancing their confidence and improving their experience.
}

\subsection{Lessons Learned}
\label{ssec:lessons-learned}

Throughout the development and the evaluation of \textit{Xavier}, we have gained valuable insights and lessons:

\textbf{\DIFdel{On the user interface design, }Presenting contexts to users in an always-on side panel makes users context-aware while minimizes excessive interruptions.}
Prior work~\cite{design-code-assistant, supernova, computational-conversational-notebook} has extensively discussed the user interface design of code assistants in computational notebooks.
For instance, the display style can be on-demand for situational contexts or always-on for continuous contexts~\cite{supernova}.
Since \textbf{DI} and \textbf{CA} are interleaved from our observation in the preliminary study (Section~\ref{sec:study-findings}) and it is uncertain when users need to view data contexts, adopting the always-on display style is a suitable design to facilitate users to remain aware of data contexts in authoring data wrangling scripts.
In comparison to displaying contexts inline, placing contexts in the side panel may help reduce disturbance while users are authoring code.
Previous code assistants like Notable~\cite{notable} and AutoProfiler~\cite{autoprofiler} adopt a similar design to reduce context switches while maintaining awareness of specific contexts like data facts or data profiles.
In our user study, with the always-on display, participants could refer to the highlighted data and the preview result at any time, with less context switch overhead.
Although the highlighted area kept changing, participants did not think the frequent change of the data view scattered their attention since they focused on the code editor while coding, which may suggest the feasibility of the always-on side panel design.

\DIFdel{
\textbf{On the code completion design, Controlling the length of suggestions are essential in scenarios with little template-based code.}
Currently, \textit{Xavier} employs various kinds of completions including column name, parameter or even transformation statement completion, in order to better adapt to different users' needs and different usage scenarios, such as fine-tuning code parameters or reusing the same operation on different data.
However, in programming tasks without specific templates like data wrangling, lengthy code completion might be difficult to fully match the user's current intent and require careful examination by users.
In our user study, participants preferred relatively short completion of \textit{Xavier} like column name or parameter completion, since it was easy to verify and aligned with their habitual way of thinking step-by-step when authoring data wrangling code.
Although many participants have adopted transformation statement completions while scripting, they still did not have overly high expectations for the accuracy of such completions.
Hence, in scenarios where template-based code is rare and user-written code is strongly correlated with their own intentions, code completion tools should prioritize performance on shorter completions for better control and verification by users.
}

\DIFadd{%
\textbf{Controlling the length of code completions can improve user experience.}
According to Section~\ref{sssec:qualitative-res}, many participants preferred shorter completions.
One of the reasons is that they believed longer completions were less accurate.
Another potential reason is that when completions did not exactly follow users' intent (e.g., \textit{Xavier} ``\textit{might suggest variations of a function}''), shorter completions can be more easily understood and verified.
Given the current limitations in model performance, it is necessary to introduce new methods to control the generation of shorter completions, making it easier for users to verify and accept the suggestions.
\CAMadd{However, current suggestions on shorter completions could be only an interim solution.}
\CAMdel{Moreover, to}\CAMadd{To} better \CAMdel{align with user intentions}\CAMadd{meet user expectations}, it would be beneficial to delve into research on models with higher performance in longer completions.
Meanwhile, it would be worthwhile for follow-up studies to develop new methods that not only foster user trust but also help assess the performance of models.
}

\DIFadd{\textbf{Improving model transparency facilitates evaluating the capability of code completion.}
Currently, \textit{Xavier} incorporates preview and highlight features to assist users in validating AI-generated code completions.
However, according to Section~\ref{sssec:qualitative-res}, these features cannot fully address the issues of transparency and trust, making participants unclear about the model's capabilities in code completion.
We suggest two possible reasons for the issues.
First, to offer straightforward code verification and mitigate information overwhelming, \textit{Xavier} did not provide explanations on what contexts were selected to complete the code.
Hence, some participants (U4, U6, U15, U16) were curious about how the code completion worked in the user study interview.
Second, users do not have a clear understanding of the performance of LLMs.
Therefore, initial few code completions may influence participants’ trust in capabilities of the tool (U8, U11, U13, U15).
To further address the transparency and trust issue, future improvements to the design of \textit{Xavier} could include presenting the contexts used for code completion to users on demand.
Additionally, quantitatively evaluating the model's performance and informing users about the model accuracy will also help them better understand the capabilities of the code completion tool.
}

\subsection{Limitations and Future Work}
\label{ssec:limitation-future}

\DIFadd{The limitations of our research can be observed in two main aspects: the functionalities and the evaluations of \textit{Xavier}.}

\subsubsection{Functionalities}
The functionalities of \textit{Xavier} can be further improved \DIFdel{according to user feedback in our user study}\DIFadd{from the three aspects}:
\DIFadd{\textbf{Data context sampling}.
To offer data context-aware code completions, \textit{Xavier} selects relevant data contexts according to the partial code being edited.
According to Section~\ref{sssec:data-context-organize}, if the amount of data contexts is large, only a sample will be collected to compute the code completions due to the limitation of LLMs.
During the development of \textit{Xavier}, we have attempted to improve the performance of code completions by adjusting different sampling sizes for data contexts, such as the different maximum number of unique values.
However, due to the lack of benchmarks, it remains unclear what the optimal amount of data contexts is, necessitating future experiments for verification.}

\textbf{Response speed}.
While using \textit{Xavier} for code authoring, users may experience delays in completion responses due to network latency, as \textit{Xavier} frequently sends requests to the LLM behind the scene.
Another reason for response delay is the relatively high computation cost, since \textit{Xavier} combines both code contexts and data contexts to generate intelligent code suggestions.
However, response time is also an essential factor affecting the user experience of code completion tools~\cite{comp-expect}.
We encourage future work to optimize the usage of both code contexts and data contexts to save the computation cost and improve response speed.

\textbf{Supported libraries}.
\textit{Xavier} uses Pandas as an exemplary data transformation library to offer data context-aware coding assistance.
To provide intelligent code completions, highlight relevant data and preview the result, \textit{Xavier} needs to parse the structure of Python Pandas code.
The parsing rules need adjustment for generalization to other programming languages and data transformation libraries.
Future work can explore a unified way like constructing intermediate domain specific language to adapt to different programming languages and libraries.

\subsubsection{Evaluations}
\DIFadd{To assess whether \textit{Xavier} facilitated locating and understanding relevant data, we defined a metric called ``context switches'' in Section~\ref{sssec:quantitative-res} to estimate the additional overhead of authoring data wrangling scripts.
These ``context switches'' assume that users pause typing data wrangling scripts while users are manually profiling datasets.
However, this metric may not sufficiently capture attention shifts of users (e.g. users might simultaneously observe the live data view and type).
As suggested by an anonymous reviewer, an interesting follow-up study could involve eye-tracking to understand how the live data view supports on-the-fly validation of code suggestions and how frequently users shift their attention between the code editor and live data view.
Such a study could provide deeper insights to further optimize the design of \textit{Xavier} and better evaluate the tool’s effectiveness.}

\section{Conclusion}

During data wrangling code authoring, users have to constantly locate and understand relevant data while writing custom scripts.
In this paper, we propose a novel coding assistance approach that prioritizes data contexts and allows users to remain aware of data contexts.
We first conducted a preliminary study to identify common patterns from a code authoring experiment, deriving three user requirements according to the semi-structured interview.
Then based on the requirements, we propose \textit{Xavier}, a computational notebook extension designed to enhance data wrangling script authoring.
\textit{Xavier} integrates both code and data contexts for data context-aware code completion, automatically highlights the most relevant data and instantly previews data transformation results based on the user's code.
\textit{Xavier} was overall appreciated by data analysts in the user study with 16 data analysts.
The coding assistance of data context awareness has the potential to be generalized to other data wrangling libraries and programming languages.
Furthermore, a long-term evaluation based on real-world datasets deserves exploration in the future work.
\begin{acks}
The research was supported by National Key R\&D Program of China (2022YFE0137800), NSFC (62402421), and HK RGC GRF grant (16210722).
The authors would like to thank Qixin Liu, Qiaosong Ying and Kaicheng Shao for helping annotate the video data of the user study.
We gratefully thank the anonymous reviewers for their valuable comments and all participants in our studies.
\end{acks}
\bibliographystyle{ACM-Reference-Format}

\end{document}